\documentclass[prd,preprint,nofootinbib,showpacs]{revtex4}
\usepackage{graphicx}
\usepackage{subfigure}

\newcommand{\be}{\begin{equation}}
\newcommand{\ee}{\end{equation}}
\newcommand{\bea}{\begin{eqnarray}}
\newcommand{\eea}{\end{eqnarray}}

\newcommand{\del}{\partial}

\renewcommand{\ap}{\alpha'}
\renewcommand{\d}[1]{\dot{#1}}
\renewcommand{\b}[1]{\bar{#1}}
\newcommand{\ket}[1]{| #1 \rangle}

\newcommand{\inner}[2]{\langle #1 | #2 \rangle}
\newcommand{\comment}[1]{{}}

\begin{document}

\preprint{hep-th/0409162}
\preprint{CALT-68-2520}
\preprint{NORDITA-2004-77}

\author{Rebecca Danos}
\email{rdanos@physics.ucla.edu}
\affiliation{Department of Physics, University of California, Los Angeles\\
Los Angeles, CA 90095}

\author{Andrew R. Frey}
\email{frey@theory.caltech.edu}
\affiliation{California Institute of Technology, 452-48\\
Pasadena, CA 91125, USA}

\author{Anupam Mazumdar}
\email{anupamm@nordita.dk}
\affiliation{NORDITA, Blegdamsvej -17\\ DK-2100, Copenhagen, Denmark}

\title{Interaction Rates in String Gas Cosmology}


\begin{abstract}
We study string interaction rates in the Brandenberger-Vafa scenario, the very
early universe cosmology of a gas of strings.  This cosmology starts with
the assumption that all spatial dimensions are compact and initially 
have string scale radii; some dimensions grow due to some thermal or 
quantum fluctuation which acts as an initial expansion velocity.
Based on simple arguments
from the low energy equations of motion and string thermodynamics, we
demonstrate that the interaction rates of strings are negligible, so the
common assumption of thermal equilibrium cannot apply. 
We also present a new analysis of the cosmological evolution of strings
on compact manifolds of large radius.  Then we 
discuss modifications that should be considered
to the usual Brandenberger-Vafa scenario.  To confirm our
simple arguments, we give a numerical calculation of the annihilation
rate of winding strings.  In calculating the rate, we also show that 
the quantum mechanics of strings in small spaces is important.
\end{abstract}

\maketitle

\section{Introduction}\label{s:intro}

In \cite{Brandenberger:1989aj}, Brandenberger and Vafa (BV) proposed a 
seemingly very natural initial condition for cosmology in string theory.
In BV cosmology, all nine spatial dimensions are compact (and toroidal
in the simplest case) and initially at the string radius.  The matter
content of the universe is provided by a Hagedorn temperature gas of
strings.  In addition to proposing a very interesting initial condition
and analyzing the thermodynamics of string at that point, however, BV 
argued that string theory in such a background provides a natural mechanism
for decompactifying up to three spatial dimensions (that is, allowing three
spatial dimensions to become macroscopic).  The BV mechanism works because
winding strings provide a negative pressure, which causes contraction
of the scale factor, as was shown explicityly 
in \cite{Tseytlin:1992xk,Tseytlin:1992ss}.  BV then gives a classical 
argument that long winding strings can only cross each other 
in three or fewer large spatial dimensions.  Therefore, since winding
strings freeze out quickly in four or more large spatial dimensions, 
the winding strings would cause recollapse of those large dimensions.

\cite{Brandenberger:1989aj} has inspired a broad literature.
One important generalization has been including branes in the early
universe gas of strings \cite{Maggiore:1998cz,Alexander:2000xv,Boehm:2002bm,
Alexander:2002gj}, and other spacetime topologies have also been
considered \cite{Easson:2001fy,Easther:2002mi}.  
Starting with \cite{Tseytlin:1992xk,Tseytlin:1992ss}, a number of authors
have examined the cosmological equations of motion appropriate to string
gases as well as brane gases \cite{Easther:2002qk,Watson:2002nx,
Watson:2003gf,Brandenberger:2003ge,Bassett:2003ck,Borunda:2003xb,
Kaya:2003vj,Campos:2003gj,
Watson:2003uw,Kaya:2004yj,Patil:2004zp,Battefeld:2004xw,Kim:2004ca,
Arapoglu:2004yf,Berndsen:2004tj}.  In particular, 
\cite{Brandenberger:2001kj,Easson:2001re} showed that interesting cosmological
dynamics happen when the expanding dimensions are still near the string
scale, and \cite{Mukherji:1996ta,Alexander:2002gj,Campos:2003ip,Campos:2004yn}
have begun to examine the effects of nontrivial backgrounds of
fields other than the metric in the low energy supergravity.

Importantly, several tests have been made of the BV mechanism for determining
the number of macroscopic dimensionality of space.  A test of the classical
string interaction rate \cite{Sakellariadou:1996vk} agreed with the 
BV argument that three or fewer large dimensions are allowed.  However,
another early test \cite{Cleaver:1995bw}
of the BV mechanism noted that long strings effectively
have a width because they can oscillate at a low energy cost, and
the effective width invalidates the classical BV mechanism.  In that event,
initial conditions determine the number of macroscopic dimensions,
which need not be three or fewer.  
Considering that winding strings have a quantum mechanical cross-sectional
width also has the same effect, which has recently been demonstrated both
in 11-dimensional M-theory \cite{Easther:2003dd} and
10-dimensional string theory \cite{Easther:2004sd}.  These tests are all
based on the fact that the winding strings will be unable to annihilate 
efficiently if their interaction rate $\Gamma$ drops below the Hubble
parameter for the expanding dimensions.

In this paper, we will also present a test of the BV mechanism in type II
string theory, along the lines of \cite{Easther:2004sd}.\footnote{In fact,
\cite{Easther:2004sd} appeared as this paper was in the final stages of
preparation.}  However, rather than follow detailed deviations from
equilibrium, we will present a general argument that winding strings will
very rapidly freeze out of a BV cosmology.  More precisely, we remind the
reader that the small-radius, high-energy density phase of a string gas is
pressureless and demonstrate that any initial conditions which lead to 
an escape of the pressureless phase also lead to freeze out of the strings.
Additionally, we briefly report numerical studies of a putative high-energy,
large-radius phase of a string gas.  In a second, largely independent
section, we calculate explicitly winding string annihilations in a BV 
cosmology, confirming that winding strings fall out of equilibrium 
immediately.  We point out that string (or particle) interactions in 
totally compact spaces generically violate conservation of energy, and we
use a perturbation theory in the time dependence of the background
to account for that violation.  This technique should be useful in future
studies of BV cosmology or any cosmology on totally compact spaces.

\section{Equilibrium Cosmology and the Dilaton}\label{s:equilib}

In this section we discuss the cosmology of strings in thermal equilibrium
in fully compact space.  In order to do so, we will review the 
thermodynamics of strings in two regimes; the first is high energy with
string scale radius, and the second is high energy and large radius.  In
each of these ``eras'' of string cosmology, we then present solutions to the
equations of motion.  Our discussion of the first (string radius) 
era is largely a review
of results presented in \cite{Cleaver:1995bw,Bassett:2003ck,Borunda:2003xb},
though we will draw some new conclusions based on the behavior of the 
dilaton.  

As is usually the case, we will assume that the spatial dimensions form a 
rectangular torus.  For simplicity, we consider two isotropic sets of
dimensions, one set which will have some initial expansion, and one set which
will remain string-scale.  For the metric, we take
\begin{equation}\label{metric}
ds^2=-dt^2+\sum_{i}^{d} R^2dx_{i}^2+\sum_{i}^{9-d} R^{\prime~2}dx_{i}^2
\,,
\end{equation}
where, for the purpose of illsutration, we have assumed that $R(t)$ is
the homogeneous scale factor for the $d$ expanding 
dimensions and rest of the spatial
dimensions, $9-d$, have scale factor $R^{\prime}(t)$.  The coordinate radii
are all $\sqrt{\ap}$.  For
convenience, we will also define
\be\label{moremetric} R=e^{\mu}\, ,\ R^{\prime}=e^{\nu}\ .\ee
The total volume of the system is given by
\begin{equation}
V=(2\pi\sqrt{\ap})^9R^{d}R^{\prime~{9-d}}\equiv (2\pi\sqrt{\ap})^9
e^{\mu d}e^{\nu (9-d)}\,.
\end{equation} 
When we discuss specific solutions to the equations of motion, we will
generally take $\nu =0$.  Further, we will henceforth take $\ap =1$.

Before we start, a few comments are in order regarding our approach to 
string thermodynamics.  First of all, we consider a thermodynamical gas of
strings on a classical (super)gravity background, as
opposed to including the gravitational variables in the
thermodynamical treatment \cite{Easther:2003dd}.  The latter approach is 
mainly important for determining the likelihood of different initial 
conditions; we take a somewhat different tack.  Another point is that we
will use the microcanonical ensemble, which is appropriate for a
totally compact universe.  The microcanonical description of string 
thermodynamics has been developed in
\cite{Deo:1989jj,Deo:1989bv,Bowick:1989us,Turok:1989bj,Deo:1991af,Deo:1992mp}.
A review of string thermodynamics,
including more recent references for microcanonical calculations,
appears in \cite{Barbon:2004dd}.

\subsection{Pressureless Phase at String-Size Radii}\label{ss:pressureless}

First we will discuss the Hagedorn phase of strings when all the
spatial dimensions have radii near the string scale, $R\sim R'\sim 1$.  

As a preliminary, we should introduce the equations of motion
for the supergravity background in the presence of a gas of strings. 
For simplicity, we will ignore the NSNS 3-form field strength and all
the RR backgrounds, as is customary.\footnote{See
\cite{Mukherji:1996ta,Campos:2003ip}
for some progress including the NSNS 3-form and 
\cite{Alexander:2002gj,Campos:2004yn} for forms in M-theory.}
It is convenient to introduce a dimensionally reduced dilaton
\be
\psi\equiv
2\Phi-d\mu -(9-d)\nu\,,\label{reddil}
\,  ,
\ee
where $\Phi$ is the dilaton of the 10D string theory.
We can now write down the equtaions of motion 
\cite{Tseytlin:1992xk}
\begin{eqnarray}
-d\dot\mu^2-(9-d)\dot\nu^2+\dot\psi^2 &=&e^{\psi}E\,,\label{constraint}\\
\ddot\mu-\dot\psi\dot\mu &=&\frac{1}{2}e^{\psi}P_{d}\,,\label{mueom}\\
\ddot\nu-\dot\psi\dot\nu &=&\frac{1}{2}e^{\psi}P_{9-d}\,,\label{nueom}\\
\ddot\psi-d\dot\mu^2-(9-d)\dot\nu^2 &=&\frac{1}{2}e^{\psi}E\, .
\label{dileom}
\end{eqnarray}
$E$ represents the total energy of the system.
The variables $P_d,P_{9-d}$ are related to the pressures $p_d,p_{9-d}$
in the respective directions by a volume rescaling, $P=pV$.  
Note that we will henceforth refer to $P$ as the pressure.  (By
referring to pressure ``in a direction,'' we strictly speaking mean the
appropriate diagonal component of the stress-energy tensor.)
We have assumed that the 10D dilaton $\Phi$ has no potential in this 
early stage of cosmology, as the dilaton must be free for winding strings
to cause the recontraction of expanding dimensions 
\cite{Brandenberger:1989aj,Tseytlin:1992xk}. This assumption, of course,
does not preclude moduli-stabilizing potentials later in cosmology.
Additionally, we have assumed that the matter action (which we can think of
as a thermodynamical free energy) is independent of the
dilaton $\psi$; relaxing this assumption changes only 
equation (\ref{dileom}) \cite{Tseytlin:1992ss}.  Note that (\ref{constraint})
implies that $\psi$ is monotonic, and we take it to be decreasing,
as is usual both to maintain perturbative string theory and prevent 
runaway solutions.  Finally, we remind the reader that there are 
corrections to the equations of motion at higher orders in $\ap$; these
corrections are negligible in our cosmology as long as all time
derivatives are smaller than unity.

Therefore, the important thermodynamical data for our purposes are the
total energy and pressure.  
The appropriate microcanonical thermodynamics for strings in
string scale, totally compact spaces have been discussed in
\cite{Brandenberger:1989aj} and subsequently in
\cite{Deo:1989jj,Deo:1989bv,Bowick:1989us,Turok:1989bj,Deo:1991af}.  
The basic thermodynamical relationships in the microcanonical ensemble
are 
\be\label{thermo}
P_d = T\frac{\del S}{\del\mu} \ ,\ \ P_{9-d}=T\frac{\del S}{\del\nu}
\ ,\ \ \frac{1}{T}= \frac{\del S}{\del E}\ ,\ee
where $S$ is the entropy.  
According to \cite{Deo:1989jj}, the microcanonical density of states at large 
energies
(when corrected to account for conservation of charge \cite{Deo:1989bv}) 
is given by\footnote{The canonical partition function is also calculated in
\cite{O'Brien:1987pn,Alvarez:1987sj}.}
\be\label{partition}
\Omega = \int_{L-i\infty}^{L+i\infty} \frac{d\beta}{2\pi i} e^{\beta E}
Z(\beta, R,  R')\ ,\ \ Z \simeq \frac{1}{E^9(\beta-\beta_0)}
\prod_{n\neq 0} \left(
\frac{\beta_0-\beta_n}{\beta-\beta_n}\right)^{g_n}\ .\ee
On notation: $\beta_0 =2\pi(2\ap)^{1/2}=\sqrt{8\pi^2}$ 
is the Hagedorn temperature for 
type II strings, the other temperatures are
\be\label{betasings}
\beta_n = \beta_0\left[ 1-\frac{1}{2}\sum_i \left(\frac{n_i}{R_i}\right)^2
\right]^{1/2}\ \textnormal{or}\ \beta_0\left[ 1-\frac{1}{2}\sum_i 
\left(n_iR_i\right)^2
\right]^{1/2}\ee
for $\sum (n/R)^2, \sum (nR)^2<2$, and $g_n$ is the multiplicity of the 
vectors $n$ with
the same $\sum (n/R)^2, \sum (nR)^2$.  The $\simeq$ indicates equality up to
an overall (nearly constant) factor.  For the radii with 
$1<R,R'<\sqrt{2}$, there are four poles, and we get
\bea
\Omega &\simeq& \frac{1}{E^9} e^{\beta_0 E}\left[1-
\frac{e^{-(\eta_1)E}}{(2d-1)!} \left((\eta_1)E\right)^{2d-1}
-\frac{e^{-(\eta_2)E}}{(2d-1)!} 
\left((\eta_2)E\right)^{2d-1}\right.\nonumber\\
&&\left. -\frac{e^{-(\eta'_{1})E}}{(17-2d)!}\left((\eta'_1)E
\right)^{17-2d}
-\frac{e^{-(\eta'_{2})E}}{(17-2d)!}\left((\eta'_2)E
\right)^{17-2d}\right]\ .
\label{density}\eea
The poles are at
\be\label{poles}
\beta_1 = \beta_0\left[ 1-\frac{1}{2R^2}\right]^{1/2}\ ,\ \
\beta_2 = \beta_0\left[ 1-\frac{R^2}{2}\right]^{1/2}\ ,\ \
g_{1,2}=2d\ ,\ee
and $\eta_i=\beta_0-\beta_i$.
The primed poles are as in (\ref{poles}) with $R\to R'$ and $g'_{1,2}=18-2d$.

Then we obtain the leading order expression for
the temperature and the pressures (see \cite{Bassett:2003ck})
\begin{eqnarray}
\frac{1}{T}&\sim&\beta_{0}-\frac{9}{E}+C_{1}E^{2d}\sum_{i=1,2}e^{-\eta_{i} E}
+C_{2}E^{17-2d}
\sum_{i=1,2}e^{-\eta_{i}^{\prime}E}\,,\\
P_{d}&\sim&C_{3}E^{2d}e^{-\eta E}\,,\\
P_{9-d}&\sim& C_{4}E^{17-2d}e^{-\eta^{\prime}E}\,,
\end{eqnarray} 
where $C_1,C_3$ are polynomials of $\eta$ and $C_3,C_4$ are functions
of $\eta^{\prime}$.  
Note that for large enough $E$, the exponential terms dominate over 
the polynomial
terms, and, as a result, the temperature remains close to the Hagedorn 
temperature,
$T\sim \beta_{0}^{-1}$ and the pressures remain vanishingly small, 
$P_{d}\sim P_{9-d}\sim 0$.  This is precisely the regime of our interest 
(because there should be a macroscopic number of strings),
so we will take the pressures to vanish.  Assuming adiabaticity, then,
the energy should be a constant (as indeed follows from conservation of
the stress tensor).
Note that technically we should change the poles once we get to $R>\sqrt{2}$;
however, at high energies, the pressures are still exponentially suppressed,
as are the radius dependent corrections to the temperature.  Therefore,
we can maintain the approximation of a pressureless gas that maintains roughly
a constant temperature and total energy during the first
phase of cosmological expansion.

We should pause to explain why the pressure nearly vanishes.  Heuristically,
the reason is that the winding strings exert a negative pressure (like cosmic
strings), and, with all radii near the string scale, 
the population of 
winding modes and momentum modes are nearly equal (because their masses are
very similar).  In fact, T-duality implies vanishing pressure.  At the 
self-dual radius, the pressures $P,\tilde P$ in the original and T-dual
variables should be equal.  However, because $\mu,\nu\to -\mu,-\nu$ under
the associated T-duality, $P=-\tilde P$, so the pressure vanishes.
Therefore, even more general gases of strings and D-branes must have a
pressureless phase near the self-dual radius (which applies to
\cite{Alexander:2000xv,Boehm:2002bm}). 

\begin{figure}[t]
\subfigure[$\mu$ vs $t$]{\label{ff:mu_t1}
\includegraphics[scale=0.75]{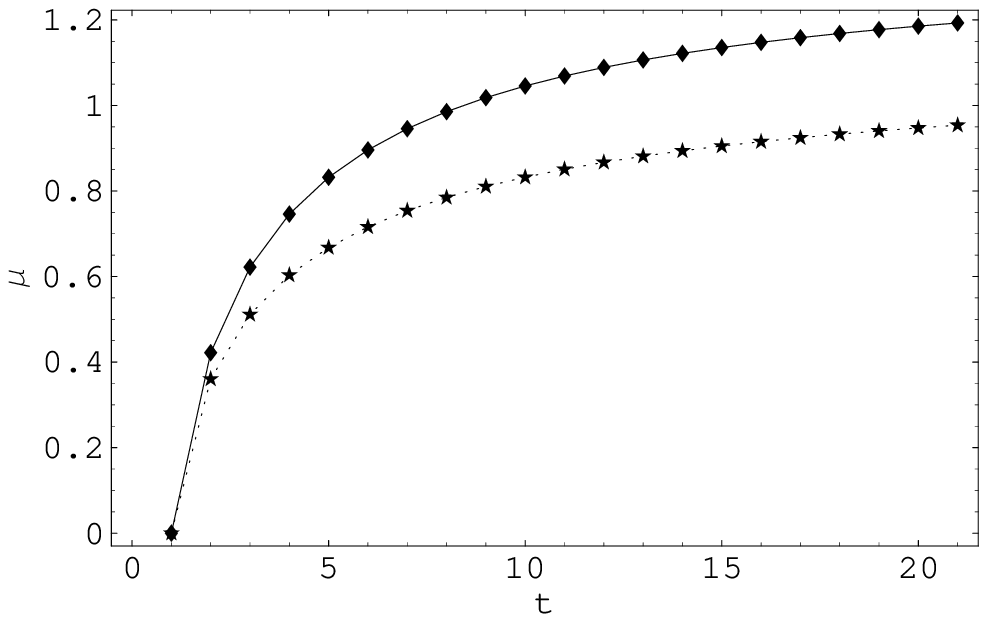}}
\subfigure[$\psi$ vs $t$]{\label{ff:psi_t1}
\includegraphics[scale=0.75]{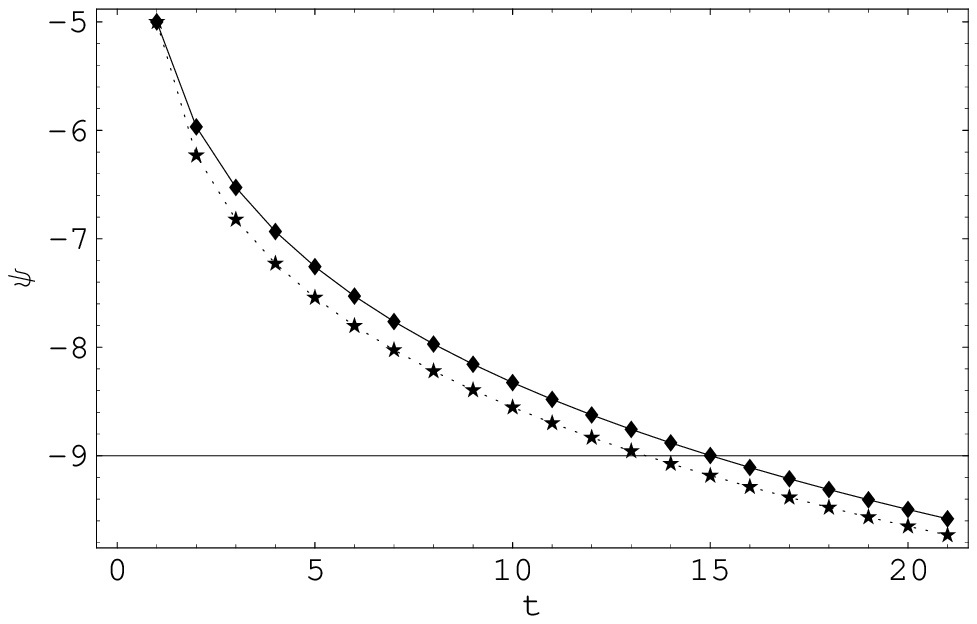}}
\caption{\label{f:era1}Cosmological evolution of the scale factor and 
dimensionally reduced dilaton for $\d\mu_0=0.7$, $\psi_0=-4$, and $E_0=100$
for 3 (solid curve, diamonds) and 9 (dashed curve, stars) 
expanding dimensions.}
\end{figure}

In this pressureless phase, the equations of motion are relatively easy
to solve, and in fact are a simple generalization of results given in
\cite{Tseytlin:1992xk,Cleaver:1995bw,Bassett:2003ck}: 
\begin{eqnarray}
e^{-\psi}&=& \frac{E_{0}}{4}t^2+Bt+\frac{B^2-dA_{d}^2-(9-d)
A^2_{9-d}}{E_{0}}\,,\\
\mu&=&\mu_{0}+\frac{A_{d}}{\alpha}\log\left[\frac{
(E_{0}t+2B-2\alpha)(B+\alpha)}{(E_{0}t+2B+2\alpha)(B-\alpha)}\right]\,,\\
\nu&=&\nu_{0}+\frac{A_{9-d}}{\alpha}\log\left[\frac{
(E_{0}t+2B-2\alpha)(B+\alpha)}{(E_{0}t+2B+2\alpha)(B-\alpha)}
\right]\,,\\
\alpha&=&\sqrt{dA_{d}^2+(9-d)A_{9-d}^2}\,,
\end{eqnarray}
where 
\be
A_{d}=\dot\mu_{0}e^{-\psi_{0}}\,,~A_{9-d}=\dot\nu_{0}e^{-\psi_{0}}\,,~
B=-\dot\psi_{0}e^{-\psi_{0}}\,.
\ee
A useful intermediate result, which follows directly from equations
(\ref{mueom},\ref{nueom}) is that both $\dot\mu,\dot\nu \propto e^\psi$.
In fact, this result is true in the pressureless phase even if
(\ref{dileom}) is modified.  For simplicity, we will consider only solutions
in which $\nu(t)=0$, so $A_{9-d}=0$.  We henceforth drop the subscript on 
$A_d$.  See figure \ref{f:era1} for plots of the scale factor and dilaton
in the distinct cases of $d=3,9$ expanding dimensions; note that the 
time evolution of $\psi$ is not that sensitive to dimension.

As discussed in \cite{Bassett:2003ck}, 
the radius $R=e^\mu$ asymptotes to the value
\be\label{asymptote}
R_\infty = e^{\mu_0}\left|\frac{B + \sqrt{d}A}{B - \sqrt{d}A}\right|^
{\frac{1}{\sqrt{d}}} \ .
\ee  
Because the radius asymptotes to a finite value, the only hope that we 
have of ``decompactification'' of the $\mu$ directions is for the string gas
to leave the pressureless phase.
As we have noted in our discussion of the thermodynamics, the gas of strings
will leave the pressureless phase when the universe is no longer near the
string radius, when $e^\mu>\b R$ for some radius $\b R$ which divides ``large''
from ``small.''  We take $\b R\approx 3$ for numerical purposes.   
In particular, for the asymptotic radius $R_\infty > \b R$, we find that
\be\label{initialdilaton}
e^{\psi_0} <   \frac{\dot{\mu_0}^2 d}{E_0} \left[\left(
\frac{(\b R e^{-\mu_0})^{\sqrt{d}} + 1}
{(\b R e^{-\mu_0})^{\sqrt{d}} -1}\right)^2 - 1\right] \ .
\ee
As far as we know, this is the first discussion of such a limit on the
initial dilaton.  Note that at large energies, the dilaton is considerably
suppressed.

Finally, we should check that our initial conditions 
avoid a Jeans instability resulting in a black 
hole, as was discussed for string thermodynamics in \cite{Atick:1988si}.  
To avoid the Jeans instability, we require 
\be\label{jeans1}
R^2 \leq \frac{1}{\kappa^2\rho}\ .
\ee
For all radii initially at $\sqrt{\ap}$, (\ref{jeans1}) is
\be\label{jeans2}
e^{\psi_0} \leq \frac{(2\pi)^9}{\kappa_0^2 E_0} \ .
\ee
The constraint (\ref{jeans2}) is somewhat less restrictive than 
(\ref{initialdilaton}) if we keep $\d\mu_0<1$ as required 
for the use of the low energy equations of motion.

\subsection{Hagedorn Strings at Large Radii}\label{ss:largeR}

Now we would like to discuss the cosmological evolution associated with
a phase of strings at high energies and large radius, which we take to 
start at $\b R>R$.  Although we will find later that the initial, pressureless
phase cannot remain in equilibrium in order to lead to this phase, 
presumably strings could equilibrate after the universe reaches a large
radius.  Therefore, these cosmological solutions may be useful in future
studies.  We know of no other reference that has considered the cosmological
evolution of this phase of strings, as other works have concentrated on
lower energy gases of strings, which are dominated by radiation (see
\cite{Bassett:2003ck,Borunda:2003xb,Easther:2004sd}).

For this later era  
\cite{Deo:1992mp} give
\be
\Omega = \frac{\beta_0}{E^9}e^{\beta_0E + (\lambda + 
a)R^d} 
\ee
for the multiple string density of states.
We include the prefactor of $1/E^9$ because momentum and winding in each
of the nine spatial directions are conserved, as discussed in 
\cite{Deo:1989bv,Deo:1992mp}.  Here $\lambda$ is a constant (which 
we take to vanish), and $a$ is constant with radius but depends
on the number of large dimensions:
\be\label{aconstant}
a= -2^{d/2} \frac{2\pi^{d/2}}{\Gamma(d/2)} \int_0^1 dx x^{d-1}\ln \left[ 1-
\frac{2}{1+\sqrt{2}}\left(1-x^2\right)^{1/2}\right]\ .\ee
The entropy is therefore
\be\label{largeentropy}
S = \ln\beta_0 - 9\ln E + \beta_0E + (\lambda + a)R^d \ .
\ee
The temperature and pressure are therefore
\bea
\frac{1}{T} &=& \frac{\beta_0 E-9}{E} \label{largetemp} \\
P &=& \left(\frac{E}{\beta_0E - 9}\right)(\lambda  + a)d e^{\mu d} \ .
\label{largepressure}
\eea

Working with the usual assumption of adiabaticity, the equations of motion
are now too difficult to solve analytically.  The main stumbling block is
that, holding the entropy (\ref{largeentropy}) fixed, the relation between
the energy and radius is transcendent.  However, it is straightforward
to solve the equations of motion numerically; we found it useful
to rewrite
\be
\mu = \frac{1}{d}\ln\left(
\frac{9\ln E - \beta_0E + S - \ln\beta_0}{\lambda + a}\right)
\ee
and solve numerically for $E$ as a function of time.

\begin{figure}[t]
\subfigure[$\mu$ vs $t$]{\label{ff:mu_t2}
\includegraphics[scale=0.75]{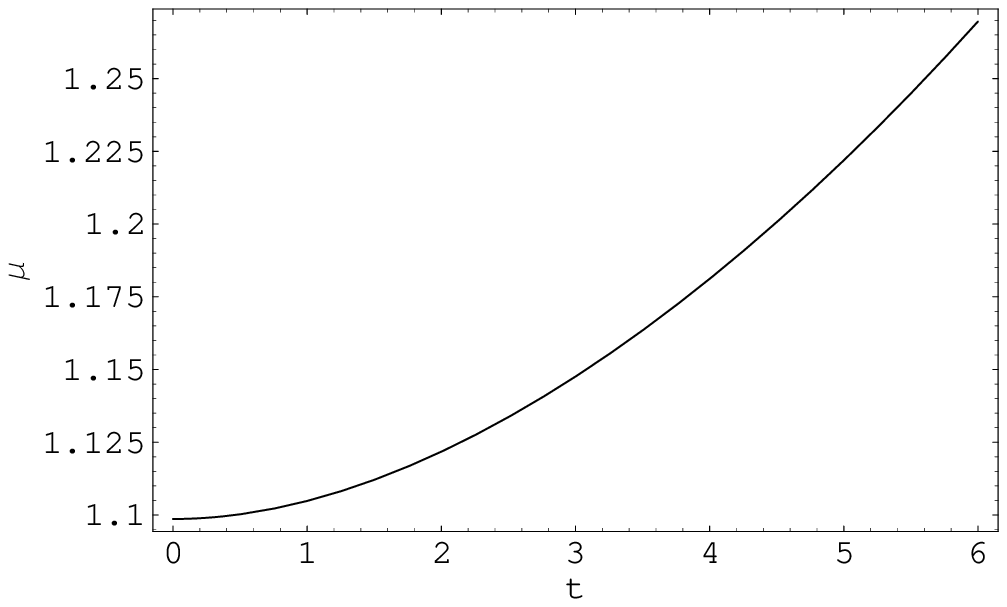}}
\subfigure[$\psi$ vs $t$]{\label{ff:psi_t2}
\includegraphics[scale=0.75]{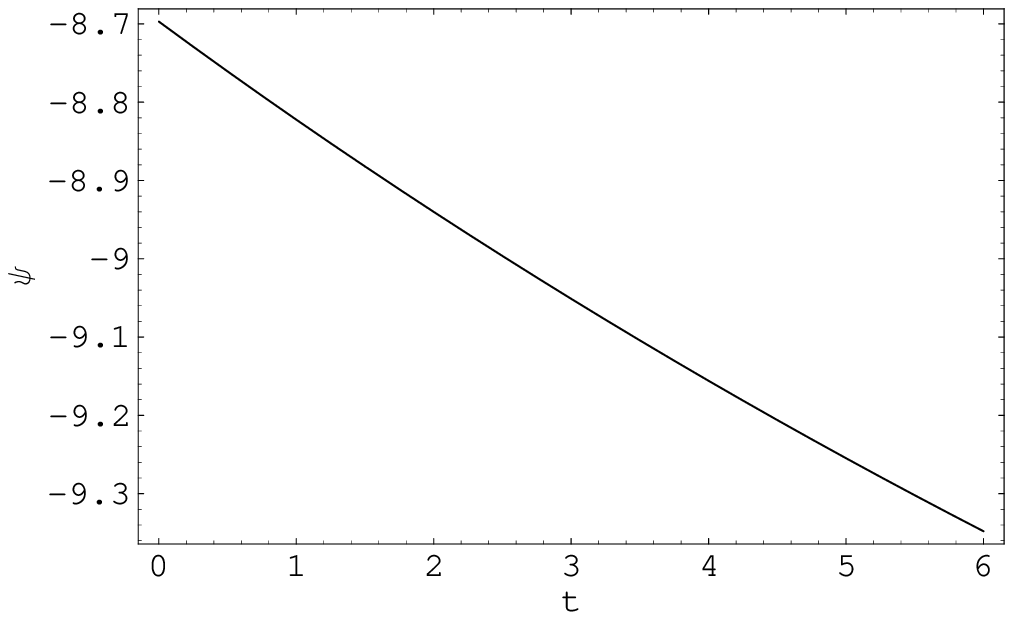}}
\caption{\label{f:era2}Evolution of the scale factor and dilaton in
the large-radius Hagedorn phase for $d=3$ expanding dimensions.  The initial
conditions are given by the pressureless phase evolution of the initial
conditions given in figure \ref{f:era1}.}
\end{figure}

We include plots of the scale factor and dilaton for this large-radius
phase in $d=3$ expanding dimensions in figure \ref{f:era2}.  
For initial conditions, we
take the values of $E$, $\dot\mu$, $\psi$, and $\dot\psi$ given
by evolving the pressureless phase to $e^\mu=\b R\equiv 3$ (although
other initial conditions might be more interesting, as indicated below).
In the pressureless phase, we took the same initial conditions as listed
in figure \ref{f:era1}.  Now the expanding dimensions can resume a rapid
expansion, while the dilaton continues its monotonic decrease. When the
energy density has decreased enough due to cosmological redshifting, this
phase should match onto a radiation dominated phase (assuming thermal
equilibrium).

\subsection{Estimating Interaction Rates from Cosmology}\label{ss:cosmorates}

Now we will consider the importance of (\ref{initialdilaton}) 
in determining the interaction rate of strings.  As a particular interaction
of interest for the BV mechanism, 
we can focus on the annihilation of winding and anti-winding
string pairs, although the key issue is whether the strings can remain
in equilibrium.  We first give a rough estimate of the interaction rate.

Up to gamma-function factors, the amplitude for any 2 string to 2 string
process (with all dimensions near the string radius) is $e^\psi$,
as we will discuss in more detail below.  Therefore,
we write the interaction rate roughly as 
\be\label{roughrate1}
\Gamma \approx \mathcal{N} \sum_{\mathbf{s}} \sum_{\vec{p}} e^{2\psi}
\ ,\ee
where $\mathcal{N}$ is the number of strings with which a given single
string can interact, and the sums are over the outgoing spin and momentum
states.  At this level of discussion, the sum over spins just gives
a constant factor of order $100$.  Also, for this 
simple estimate, we can approximate 
the sum over outgoing momentum states as a constant factor; for dimensions
near the string radius, corresponding to the pressureless phase, this
momentum sum factor is order unity.

Since the classical BV mechanism distinguishes the annihilation rates of
winding strings in $d\leq 3$ and $d>3$ large dimensions, we comment here
on the effect of varying $d$ on (\ref{roughrate1}), the quantum mechanical
annihilation rate.  The dimensionality enters into $\Gamma$ through the
time evolution of $\psi$ (and also somewhat in the constraint on its initial
value (\ref{initialdilaton})); the higher $d$, the faster $\psi$ decreases.
This effect tends to enact the BV mechanism, as strings interact less 
rapidly with more growing dimensions.  The other factor through which $d$
affects $\Gamma$ is the sum over outgoing momenta.  Quite simply, for a 
given energy, the phase space of outgoing momenta is larger in a greater number
of large dimensions.  This tends to counterbalance the variation of the
dilaton, but it is a small effect when all dimensions are near the string
radius.

The number of strings available for scattering is determined by 
thermodynamics.  For the pressureless phase, the number of strings
of charge $q$ and energy between $\epsilon$ and $\epsilon +d\epsilon$ in
a gas of strings of total energy $E$ is \cite{Deo:1989bv}
\be\label{distro}
\mathcal{N}(\epsilon,q,E) = \left(\frac{4\pi}{\beta_0}\right)^9
\frac{1}{\epsilon} \left(\frac{E}{\epsilon(E-\epsilon)}\right)
\exp\left[-\frac{E}{4\epsilon (E-\epsilon)} q^T A^{-1}q\right]\ .\ee
Here the matrix $A$ is 
\be\label{matrixA}
A = \frac{\beta_0}{(4\pi)^2} \textnormal{diag}(
\underbrace{1/R,\cdots}_{d},\underbrace{1/R',\cdots}_{9-d},
\underbrace{R,\cdots}_d,\underbrace{R',\cdots}_{9-d})\ .\ee
This takes into account the contribution of winding and momentum, as well
as oscillator modes, to the energy of the string.
The total number of strings in the gas, given by summing over charge and
integrating over $\epsilon$ is 
$\mathcal{N}=\int d\epsilon \sum_q \mathcal{N}(\epsilon, q,E)\approx \ln E$.  
Actually, 
restricting to strings of a certain charge or range of energies gives a 
considerably smaller number, so we can take $\mathcal{N}\lesssim \ln E$ as
an upper bound in (\ref{roughrate1}).

If we substitute in the bound (\ref{initialdilaton}), our estimate of the
interaction rate becomes 
\be\label{roughrate2}
\Gamma\lesssim 100 \left(\frac{\d\mu_0^2 d}{E_0}\right)^2\ln E_0\ee
during the entire pressureless phase.  We have used the facts that the
total energy in the string gas is constant during the pressureless phase
and that the dilaton is monotonic decreasing.
Even for $E_0\approx 100$, which only leads to about 4 strings in equilibrium,
we have $\Gamma <\d\mu_0$.  For larger energies, the interaction
rate will only decrease.

Therefore, it seems inevitable that the string gas will fall out of 
equilibrium almost immediately as the universe begins expanding in the
pressureless phase.  
This means that, for any $d$, winding strings will freeze out and 
come to dominate the
energy density of the compact universe and cause its recollapse.  Contrast
this case to the original BV mechanism, which claims that winding
strings can annihilate effeciently for $d\leq 3$.\footnote{Of course,
in more complicated compactifications, such as orbifolds, winding strings
can decay out of equilibrium.  Then the question is whether they decay
too rapidly to produce the BV mechanism \cite{Easson:2001fy,Easther:2002mi}.}
The reader might wonder if the string gas could come into equilibrium as
$\dot \mu$ decreases in time; however, if we assume pressureless evolution,
$\dot\mu\propto e^\psi$ while $\Gamma\propto e^{2\psi}$ decreases more 
rapidly.

As a result, any study of the standard BV
scenario should focus on the non-equilibrium thermodynamica
and cosmology of string gases,
as partially discussed in \cite{Easther:2003dd,Easther:2004sd}
(these references included the non-equilibrium Boltzmann equation for 
strings but used the equilibrium cosmological evolution).
There are a few possibilities to avoid this 
conclusion; however, they all modify the usual BV scenario.
A simple idea would be to consider lower energy string gases (ie, 
$\beta\gg \beta_0$), as in \cite{Patil:2004zp}, but winding states would be
populated only infrequently, so it might be difficult for winding strings to
stabilize the radii of any of the dimensions.
One obvious possibility is to take $\d\mu_0 \gg 1$, which would require
understanding the full time-dependent worldsheet CFT.  Unfortunately, this
is beyond our capabilities, and in fact such a CFT may not have a 
straightforward geometric interpretation.  Another possibility would be
to consider the monotonically increasing solution for the dilaton, which 
could lead to a strong gravity regime and large time derivatives.
Alternately, heterotic strings have qualitatively different thermodynamics
due to additional low-energy states near the string radius, so they may
evade (\ref{initialdilaton}).  However, at high enough energies, even heterotic
compactifications should have a phase similar to the pressureless phase
described above.

Another attractive option is the introduction of winding branes in the 
gas of strings, as has been widely considered \cite{Alexander:2000xv,
Easson:2001fy,Boehm:2002bm,Alexander:2002gj}.  
Gases of D-branes might avoid our constraints because the branes couple to
the dilaton and therefore modify equation (\ref{dileom}).  However, the
other equations of motion remain the same
\cite{Tseytlin:1992ss}, and, because brane gases are 
indeed T-duality invariant \cite{Boehm:2002bm}, the early phase of 
cosmology should still be pressureless.  Therefore, $\d\mu\propto e^\psi$
as before.  As long as $\psi$ decreases quickly enough, $\mu$ will still
asymptote to a finite value, leading to a constraint on the initial value
of the dilaton, as in (\ref{initialdilaton}).  In a perturbative
model, in fact, we expect D-brane states to be populated very little, so
branes will not have much effect on the time evolution of $\psi$. 
Therefore, it seems reasonable that (\ref{initialdilaton}) will hold
approximately even if branes are included in the thermal gas of strings.
See \cite{Brandenberger:2001kj,Easson:2001re,Watson:2002nx,
Brandenberger:2003ge,Watson:2003gf,Kaya:2003vj,Campos:2003gj,Kaya:2004yj,
Kim:2004ca,Battefeld:2004xw} 
for some work relevant to time evolution in brane gases.
Similarly, allowing the shape moduli of the spatial torus and the
various form fields of supergravity to be nontrivial may alter the qualitative
behavior of the dilaton, such as by providing a potential 
(for a review of flux compactifications with stabilized dilaton,
see \cite{Frey:2003tf}).  
Some work including other degrees of freedom in the context of string and
brane gas cosmology appears in 
\cite{Mukherji:1996ta,Alexander:2002gj,Campos:2003ip,Watson:2003uw}.
We must caution that including a potential for the dilaton appears to 
prevent radius stabilization by winding strings, however
\cite{Berndsen:2004tj}.

As a final alternative, we propose a stochastic version of the BV scenario.
In this case, we could imagine that the dilaton is large enough to maintain
thermal equilibrium of strings (so that the winding modes can annihilate),
but the expanding dimensions asymptote to a small radius.  Then, if some
of these dimensions were to start expanding again due to some thermal 
or quantum fluctuation, $\mu_0 >0$, so the final radius (\ref{asymptote})
can grow.  Eventually, though a random walk, some dimensions could grow large
while maintaining thermal equilibrium in the string gas.  There are some
difficulties with this proposal, however.  One is understanding the
quantum mechanics and thermodynamics of the supergravity background.  Another
is that the dilaton will still decrease monotonically in time (except 
perhaps for thermal and quantum fluctuations), so the string gas might
fall out of equilibrium in any event.

All of the above proposals are candidates to allow the string gas to remain
in thermal equilibrium and the winding modes to annihilate. However, it
is far from obvious that any of them discriminate among the number of
expanding dimensions $d$.  That is, in any of the above proposals, the 
values of $d$ which allow strings to remain in thermal equilibrium until
all the winding strings annihilate (statistically speaking) seem likely
to depend on initial conditions rather than $d$ itself.  Perhaps the most
encouraging of our proposals in that regard is the inclusion of other 
form fields, which have varying rank, which could influence dimensionality
to some extent.

%%%%%%%%%%%%%%%%%%%%%%%%%%%%%%%%%%%%%%%%%%%%%%%%%%%%%%%%%%%%%%%%%%%%
\section{Interactions in Totally Compact Spaces}\label{s:rates}

In this section, we will calculate the annihilation rate for a winding
and anti-winding string pair in order to confirm that the strings cannot
stay in equilibrium in the pressureless phase.  In so doing, we point
out a problem with the naive use of string perturbation theory in the
BV scenario and point out a resolution that could be useful in future
tests of the BV mechanism.

As we have discussed, a key feature of the string-gas and brane-gas 
cosmologies is that all the spatial dimensions are compact; in fact, much of
the interesting dynamics that determines how many dimensions can expand
to macroscopic size occurs when all the dimensions are within an order
of magnitude of the string scale (loitering, for example, occurs at 
roughly 2 to 3 times the string length 
\cite{Brandenberger:2001kj,Easson:2001re}).
Because all the dimensions are compact and in fact at a small scale, 
momentum must be considered as being quantized.   When the large dimensions
become much larger than string scale, we are justified in ignoring the 
momentum quantization because the energy spacing will be much smaller than
other energy scales.  However, since we are interested in string scale
dimensions, the energy spacing of momentum modes is nearly as large as
the string mass or even winding modes, we have to quantize momenta, even
those in the ``large'' dimensions.

Quantization of all the momenta puts interesting constraints on the 
interactions of strings, and, in particular, at generic compactification
radii, interactions cannot conserve energy.  In other words, superstrings
on a time-independent $T^9$ (or presumably any other compact space) cannot 
interact.  To see this in more detail, consider some simple 
string amplitudes:
\begin{enumerate}
\item Supergravity mode interaction in which two strings with momentum in
the $x^1$ direction scatter into the $x^2$ direction.  In the simplest 
case, the strings have no other momenta, so the initial strings have
$n_1,-n_1$ momentum numbers and the final strings have $n_2,-n_2$.  
Conservation of energy would then require $2n_1/R_1 = 2n_2/R_2$,
which is clearly impossible at generic radius since the $n_i\in\mathbf{Z}$.
\item Oscillator mode decay to momentum modes.  
If the final state strings have momenta in one direction only, energy
conservation would require $N=2n/R$, which is violated at nonintegral radius.
\item Winding mode annihilation, which is our main focus.  In the simplest
case, there
are two strings with opposite winding number $w,-w$ in the $x^1$ direction,
which interact to produce, for example, momentum numbers $n,-n$ in the
$x^2$ direction.  Energy conservation would be $2wR_1=2n/R_2$, which is
again clearly impossible for generic radii.
\end{enumerate}
In fact, even at radii (for example, $R=1$) 
where the simplest case interactions listed above
are allowed energetically, there are string states with multiple nonzero
quantum numbers that cannot interact without violating energy conservation.
Essentially, the problem is that at any given radius only a small subset
of string interactions might be allowed by conservation of energy.

There are several possible resolutions to this difficulty.  One is that the
radii are quantum mechanical variables, so that there is some probability 
of being at a radius where any given interaction is allowed kinematically.
Second, the more energetic states might have a width, or uncertainty in 
their energy, due precisely to the interactions in question.  The logic is
somewhat circular, though, as the width allows the interactions which give
rise to it.
A final solution to the dilemma
is to remember that energy is not strictly conserved in an expanding universe.
For a scale factor given by $R=e^\mu$, we expect that there will be 
energy nonconservation of order $\d\mu$ in each interaction.  Since the
smallest energy quantum is $1/R$, we expect that all interactions would be
allowed when the radius of the universe is longer than the Hubble time.
We will focus on this third solution in this section, as it is in some ways
the most straightforward to understand, and we discuss a formalism for 
dealing with compactification radii quantum mechanically in the appendix.
In all of these approaches, we would assume that, although individual 
interactions would not conserve energy, the total thermodynamical energy would
be conserved (that is, the nonconservation of energy would average to zero
over many interactions).

\subsection{Time Dependence and Energy Nonconservation}\label{ss:timedep}

Because we have been careful to ensure that all time derivatives in the
supergravity background are smaller than string scale, we can use a very
simple approach to include the expansion of the universe in the string
interactions.  Namely, rather than incorporate time dependence directly
into string perturbation theory, we develop an effective quantum 
mechanics based on string perturbation theory and then consider the
time dependence of the effective quantum mechanics.
We will focus on winding mode annihilation to momentum modes, 
$W\b W\to N\b N$, but our
approach can easily be generalized to any interaction.

In its simplest form, we can take a two-state quantum mechanics,
with the winding/anti-winding string pair ($W\b W$) as one state and
the momentum string pair ($N\b N$) as the other.  We derive the
Hamiltonian as follows.   The
diagonal, mass part comes directly from the perturbative string spectrum
on the fully compact space, and we do not need to discuss it further.
Including an interaction piece, the Hamiltonian is
\be H = \left[\begin{array}{cc}2|\vec{w}|e^{\mu} & V_I \\
    V_I & 2|\vec{n}|e^{-\mu} \end{array}\right]\ , \ee
assuming that the momentum and winding vectors $\vec n$ and $\vec w$ are 
completely in
the expanding directions (the factors of 2 in the diagonal elements are
because each state represents a pair of strings).  To find $V_I$, we
compare to the corresponding amplitude 
$2\pi\delta(E_f-E_i)\mathcal{A}$ for 
$W\b W\to N\b N$ in string perturbation theory (which we discuss in more
detail in section \ref{ss:stringamp}).
Standard string perturbation theory assumes a time independent 
background and allows an infinite time for the
strings to interact.  In quantum mechanical perturbation theory in this
case,
we find the same amplitude
\bea
\inner{\vec{n},-\vec{n}}{\vec{w},-\vec{w}} &=& -i\int dt^{\prime} 
e^{i(E_f - E_i)t^{\prime}}V_I \nonumber \\
&=& -iV_I2\pi\delta(E_i - E_f)
\eea
for a time independent $V_I$.  Taking into account the relativistic 
normalization of states in string perturbation theory, 
\be
V_I = -\frac{\mathcal{A}}{\sqrt{\prod_{i=1}^4(2E_i)}}\ ,
\ee
where the $E_i$ are the energies of the incoming and outgoing strings.
Extracting the dependence on supergravity fields,
$V_I=Ce^\psi$, where
\be
C = \frac{e^{-\psi}\mathcal{A}}{\sqrt{(2E_1)(2E_2)(2E_3)(2E_4)}}
\label{Cconstant} \ee
is a constant factor.
One advantage of this effective quantum mechanics approach is that the 
string amplitude automatically includes single string intermediate states,
as we will discuss in more detail below.
To include the evolution of the universe at any given time $t_0$, we
can expand $H$ to first order in time derivatives,
$H=H_0+H_t$.  Then we can treat
the first order term as a perturbation around the zeroth order term.

We should therefore first find the eigenstates of the time independent
Hamiltonian, including the interactions $V_I$.  In this way, we avoid
confusing state oscillation of the purely time independent quantum 
mechanics in the annihilation rate, which should be due purely to time
dependence in the Hamiltonian.  Using standard formulae from quantum
mechanics (see, for example, \cite{Sakurai:1994}), 
the eigenstates
are given by 
\bea
\ket{\vec w,-\vec w} &=& 
\ket{\vec w,-\vec w}^0 + \frac{V_I}{E_w^0 - E_n^0}\ket{\vec n,-\vec n}^0
\nonumber \\
\ket{\vec n,-\vec n} &=& \ket{\vec n,-\vec n}^0 + \frac{V_I}{E_n^0 - E_w^0}
\ket{\vec w,-\vec w}^0
\ ,
\eea
where the energy eigenvalues are just the unperturbed ones $E^0_w,E^0_n$.
We have treated $V_I$ as a perturbation because we are working in
a regime in which string perturbation theory is valid ($\psi<0$).

The lowest order transition amplitude for winding annihilation is
(again, see \cite{Sakurai:1994}, for example)
\bea
c^{(1)} &=&-i\left(Ce^{\psi}\dot{\psi} + 
\frac{1}{\omega}
2Ce^{\psi}w\dot{\mu}e^{\mu} \right. \nonumber \\
&&\left. + 
\frac{1}{\omega}2Ce^{\psi}ne^{-\mu}\dot{\mu}\right)
\left[\frac{te^{i\omega t}}
{i\omega} - \frac{e^{i\omega t} - 1}{(i\omega)^2}\right] 
\eea
with
\be
\omega = E_f - E_i = \frac{2n}{R} - 2wR \ .
\ee
Here, $n=|\vec n|$ and $w=|\vec w|$.

Next we note that $\d\mu t, \d\psi t < 1$ in order to ensure that we can 
treat the time dependence as a perturbation.  In fact,  this is consistent
with assuming $t\sim 1$ for the interaction time of strings, as the
supergravity approximation requires that all time derivatives be less than
unity.  Then we can take
$\omega t$ to be small, so  the probability is therefore
\bea
P &=& |Ce^{\psi}|^2\left[\frac{\dot{\psi}^2}{4} + 
\frac{w\dot{\mu}e^{\mu}\dot{\psi}}{\omega} +
\frac{\dot{\mu}ne^{-\mu}\dot{\psi}}{\omega} +\right.\nonumber \\
&& \left. \frac{w^2\dot{\mu}^2e^{2\mu}}{\omega^2} +
\frac{2wn\dot{\mu}^2}{\omega^2} + 
\frac{\dot{\mu}^2n^2e^{-2\mu}}{\omega^2}\right]t^4 \ .
\label{prob2}\eea

The total annihilation rate for a winding string ($W\b W\to N\b N$), 
as discussed in 
section \ref{ss:cosmorates} following equation (\ref{roughrate1}),
is 
\be\label{totalrate1}
\Gamma = \mathcal{N} \sum_{\mathbf{s}} \sum_{\vec n} P(\vec w,\vec n)
\ ,\ee
with notation as in \ref{ss:cosmorates}.  
Since we will ignore string polarizations for simplicity, the sum over
outgoing spins reduces to an overall constant.  In fact, conservation of
angular momentum correlates the two outgoing spins, so the overall factor will 
be the number of massless superstring spins, $256$, times a symmetry
factor of $1/2$.
The sum over outgoing momenta
should be within the (larger) expanding dimensions, as that is the type
of final state that interests us.  Before we present the results of a 
detailed computation, we will discuss the string perturbation theory
amplitude $\mathcal{A}$ that we used to define the effective quantum
mechanics.

One note: the reader may wonder why we did not take advantage of the optical
theorem to calculate the total interaction rate (up to the factor of 
$\mathcal{N}$), as in \cite{Polchinski:1988cn,Easther:2003dd,Jackson:2004zg}.
The simple reason is that we have no simple way to incorporate the time
dependence of the background into the appropriate calculation in string
perturbation theory, unlike the direct calculation that we present.

\subsection{Bosonic String Winding Annihilation Amplitude}\label{ss:stringamp}

As we discussed above, we need the string perturbation theory interaction
amplitude.
We carry out a somewhat rough analysis of the interaction, 
ignoring the 
contribution to the amplitude from string polarizations.  
This approximation allows us to replace the sum over outgoing spins
by an overall factor (and the average over incoming spins by unity) 
in the interaction rate (\ref{totalrate1}).  Therefore, we use
the bosonic string amplitude for nonpolarized strings (the tachyon 
mode if the winding and compact momenta vanish) but with the appropriate 
mass-shell for whatever polarized superstring
modes we are considering.  An analogous superstring calculation for 
tachyonic winding strings was discussed in \cite{Jackson:2004zg} using
the optical theorem; we calculate the amplitude directly at tree level
for use in the effective quantum mechanics as discussed above.
Our presentation is a generalization of the answer to a textbook problem 
from \cite{Polchinski:1998rq}.

In calculating the annihilation amplitude, we find it convenient to 
work with a unit metric and coordinate radii equal to the proper radii
of the spatial dimensions, 
but the amplitude is diffeomorphism invariant and can
therefore be used directly with the conventions in the text.
The vertex 
operators of the four interacting strings are all 
\be\label{vertex}
\frac{g_c}{\sqrt{V}} :\exp \left[ i k_L\cdot X_L +ik_R\cdot X_R\right]:\ ,
\ee
where the closed string coupling 
\be\label{gc}
g_c = \frac{1}{\sqrt{2}} (2\pi)^{5/2} \ap{}^2 e^\phi\ ,\ee
$\phi$ is the 10D dilaton, and $V$ is the total volume of the spatial 
dimensions (the product of $2\pi R$ over all dimensions).  
The amplitude is just the expectation value
of four vertex operators, and it has a prefactor (besides the vertex 
operator normalization) of $i(8\pi/g_c^2\ap)
(2\pi) \delta(\sum k^0) V \delta_n  \delta_w$.  
Here $\delta(\sum k^0)$ conserves energy
and $\delta_{n,w}$ conserve compact momentum and winding (ie, these are
products of Kronecker delta symbols).  There is also
an overall sign (the cocycle) which we will ignore as we have only a single
amplitude.  Thus, the amplitude becomes
\be\label{amp1}
\mathcal{A} = i2(2\pi)^{-3} e^{\psi} (2\pi) \delta^{10-d}(k)
\delta_n \delta_w F(k_L,k_R)\ ,\ee
with $\psi$ as before.  Here
$F$ is some function of the momenta that we need to determine,  and we are
again setting $\ap =1$.

As it turns out, we are interested in annihilation of winding, so the two
outgoing strings will have zero winding, and the two incoming strings will
have opposite winding $\vec w\equiv \vec w_1=-\vec w_2$.  
In addition,  we will take the simplest case in which the winding
strings have no momentum.  The compact momenta are
\bea
\vec k_{L3}=\vec k_{R3} =-\vec k_{L4}=-\vec k_{R4}
&=& \overrightarrow{\left(\frac{n}{R}\right)}
\nonumber\\
\vec k_{L1}= -\vec k_{R1}=-\vec k_{L2} = \vec k_{R2} &=&
\overrightarrow{\left(wR\right)}
\label{momenta}\eea
where $\overrightarrow{(n/R)}$ is the vector with components $n_i/R_i$
($R_i$ is the physical radius in the $i$th direction) and similarly for
$\overrightarrow{(wR)}$.  Additionally, $k^0_I$ represents the 
\textit{incoming} energy of the $I$th string, which is negative for the
outgoing momentum strings, while $E_I$ is the physical energy of the $I$th
string, which is always positive.

In this case, the momentum dependence becomes
\be\label{Famp}
F = \int d^2z |z|^{ E_1 E_4} |1-z|^{ E_2 E_4}
\left( \frac{z}{\bar z} \right)^{\vec n\cdot \vec w/2}
\left( \frac{1-\bar z}{1-z} \right)^{\vec n \cdot \vec w/2}\ee
integrated over the sphere.  Note that $E_{3,4}$ appear asymmetrically
due to the way in which we have located the vertex operators on the sphere;
however, the end result will be symmetric, as it must due to $SL(2,\mathbf{C}$
invariance on the worldsheet.
To do this integral, we can
integrate by parts $\vec n\cdot \vec w$ on each of $\bar z$.  Then we
some prefactors 
along with the usual integration needed for the Virasoro-Shapiro 
amplitude.  As it turns out, the $\Gamma$ functions from the integral
can absorb the prefactors,  giving us
\bea
F&=&F_1 F_2 F_3\nonumber\\
F_1&=&\frac{\Gamma\left( 1+\frac{1}{2}(E_1 E_4-\vec n\cdot \vec w) 
\right)}{\Gamma\left( -\frac{1}{2}(
E_1 E_4+\vec n\cdot \vec w)\right)}
=\frac{\Gamma\left( 1+\frac{1}{2}(nw-\vec n\cdot \vec w) 
\right)}{\Gamma\left( -\frac{1}{2}(
nw+\vec n\cdot \vec w)\right)} 
\nonumber\\
F_2&=&\frac{\Gamma\left( 1+\frac{1}{2}(E_2 E_4+\vec n\cdot \vec w) 
\right)}{\Gamma\left( -\frac{1}{2}(
E_2 E_4-\vec n\cdot \vec w)\right)}
=\frac{\Gamma\left( 1+\frac{1}{2}(nw+\vec n\cdot \vec w) 
\right)}{\Gamma\left( -\frac{1}{2}(
nw-\vec n\cdot \vec w)\right)}\nonumber\\
F_3&=& \frac{\Gamma\left(-1-\frac{1}{2} (E_1+E_2) E_4\right)}{\Gamma\left(
2+\frac{1}{2} (E_1+E_2)E_4
\right)}
=\frac{\Gamma\left(-1-nw\right)}{\Gamma\left(
2+nw
\right)}\ .
\label{Famp2}\eea
The second expression for the $F_i$ arise from
using the mass-shell condition for the strings,
which should be of the form (in the general case)
\be\label{massshell}
E^2 = 4N +\overrightarrow{\left(\frac{n}{R}\right)}^2
+\overrightarrow{(wR)}^2 \ .\ee
$N$ is an integer that describes the oscillator excitation of the string.
For bosonic string tachyon vertex operators, such as we are using, we should
take $N=-1$; however, we want to simulate superstrings at the massless
level, so we take $N=0$.  We have therefore taken $E_{1,2}=wR$ and 
$E_{3,4}=n/R$.  

There is one additional subtlety to discuss.  By unitarity, the string
amplitude contains poles corresponding to single string intermediate states.
(This fact is actually very useful for our quantum mechanics
because it means we do not need to use second order perturbation theory
to include all interactions at the same order in $e^\psi$.  \textit{All} the 
tree-level annihilation processes are included in this single sphere 
amplitude.)  These poles appear through the gamma function in the numerator
of $F_3$, which has a pole whenever $nw$ is an integer.  Indeed, if we
rewrite $nw$ in terms of the center-of-mass energy (Mandelstam variable)
$nw=s/4=(E_1+E_2)^2=(E_3+E_4)^2$, then the poles are at
$s=-4,0,4,8,\cdots$, just the bosonic closed string oscillator spectrum.

However, since we are interested in computing the actual interaction
rates, we must include the widths of the poles.  Although we do not know the
appropriate formulation of resonances in the full time-dependent formalism,
we can, at our level of approximation, use the traditional Breit-Wigner
resonance, replacing $s\to s+im\Gamma_w$ at the pole given by $s=m^2$,
where $\Gamma_w$ is the width of the resonance (see a quantum field theory
text, such as \cite{Peskin:1995ev}).  To encompass all the 
poles at once, we use $s\to s+i\sqrt{s}\Gamma_w$, which is correct at 
each pole.  To calculate the width, we ignore string polarizations again,
so the width becomes
\be\label{width1}
\Gamma_w = 128 \sum_{\vec n} \frac{1}{2\sqrt{s}} \frac{1}{s}
\frac{8}{(2\pi)^2}e^\psi
\delta\left(\sqrt{s}-2n/R\right)\ .\ee
The delta function conserves energy, and $\sqrt{s}$ is the energy of the 
intermediate oscillator resonance.  Due to momentum conservation, the
sum over outgoing momenta is a single sum in the $d$ expanding directions,
which we approximate by an integral:
\be
\sum_{\vec n} \delta\left(\sqrt{s}-2n/R\right) \approx
R^d \int d^d\vec p \, \delta\left(\sqrt{s}-2|\vec p|\right)
= \frac{R^d}{2} \left(\frac{\sqrt{s}}{2}\right)^{d-1} 
\frac{(2\pi)^{(d-1)/2}}{\Gamma( (d-1)/2)}\ .\ee
Therefore, the width is
\be\label{width2}
\Gamma_w = 64 R^d e^\psi \left(\frac{s}{4}\right)^{(d-4)/2} 
\frac{(2\pi)^{(d-5)/2}}{\Gamma((d-1)/2)}\ ,\ee
leaving us 
\be\label{Famp3}
F_3 = \frac{\Gamma\left(-1-nw -i\sqrt{nw}\Gamma_w/2 \right)}{\Gamma\left(
2+nw \right)}\ ,\ee
with $s$ replaced by $4nw$ in $\Gamma_w$.

\subsection{Results of Calculation}\label{ss:results}

\begin{figure}[t]
\includegraphics{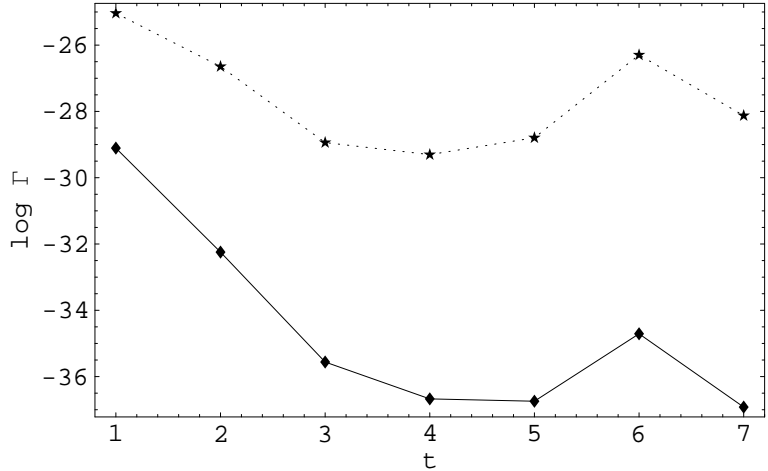}
\caption{\label{f:rate} The interaction rate for strings in the pressureless
phase for $d=3$ expanding dimensions.  The initial conditions are those
given in figure \ref{f:era1}.  The lower, solid curve has $\mathcal{N}$
calculated for $w=-1$, and the dashed curve takes $\mathcal{N}=\ln E$.}
\end{figure}

We now briefly present the results of calculating the total annihilation
rate (\ref{totalrate1}) in the pressureless phase of the cosmology.  
For the purposes of the calculation, we first
take $\mathcal{N} = \mathcal{N}(\epsilon=wR, q=\vec w, E)\Delta\epsilon$,
where we take $\Delta\epsilon =1$ in string units.  Thus, we are calculating
the annihilation rate purely for strings with $\vec w$ units of winding
and no extra oscillators.  Then, as an upper bound on the annihilation
rate, we take $\mathcal{N}=\ln E$, which is the total number of strings
in the thermodynamical gas.  We see that, in either case, the interaction
rate is small compared to the Hubble constant $\dot\mu$, so the gas
must fall out of equilibrium.  We do the momenum 
sum $\sum_{\vec n}$ by brute force, requiring at all times that the
frequency $\omega = 2n/R -2wR<1$ (since larger values are suppressed).  
For our example calculation, we take $\vec w=(1,0,0\cdots 0)$, and 
$R\leq 3$ in the first era, so we take each component of $\vec n$ between
$-10$ and $10$ for the sum over momenta.  The logarithm of the interaction
rate is plotted in figure \ref{f:rate}.  The initial conditions are
the same as those used in plotting figure \ref{f:era1}.  Note that the 
gamma functions give an additional suppression of the interaction rate.

The interaction rate is very small compared to the Hubble parameter (which 
is order unity as an initial condition).  We have carried out this calculation
also for $d=2,4$ expanding dimensions and found no qualitative difference.
There is one interesting feature of these interaction rates: a slight increase
at later times.  This temporary increase is due to an increase in phase
space available to the outgoing modes as some of the dimensions expand.

\section{Summary}\label{s:summary}

In summary, we have made two arguments regarding the usual Brandenberger-Vafa
proposal for the cosmology of string gases on totally compact spaces.  

In the first part of the paper, we analyzed two possible states of a string 
gas at high energies, reviewing their thermodynamics and solving for the
cosmological evolution of the universe in those phases.  The later phase
had not been previously studied in the literature.  From the time evolution
of the dilaton and scale factor, we were able to give a heuristic argument
that there are not initial conditions in the small-radius, pressureless 
phase that both grow into the large-radius phase and maintain thermal
equilibrium of the string gas.  We also argued that more general brane
gases would likely have a similar constraint and gave several possible
scenarios that could avoid this conclusion, all of which modify the usual
Brandenberger-Vafa scenario.  One key point is that our conclusion that
strings fall out of thermal equilibrium is largely independent of the 
number of expanding dimensions.  Therefore, it would seem like any possible
modification to the Brandenberger-Vafa scenario would not select a favored
number of large dimensions, as was originally proposed in 
\cite{Brandenberger:1989aj}.  

In the second part of this paper, we calculated the approximate 
annihilation rate for winding strings in different numbers of expanding
dimensions.  This calculation confirmed that the strings cannot remain in
thermal equilibrium in the small-radius phase of the string gas.  
In that section of the paper, we also argued that quantization of string
momenta is important during the small-radius phase.  Specifically, 
due to quantization of string winding and momenta, string interactions will
generically violate conservation of energy (and would thus seemingly be
forbidden).  We demonstrated that the time dependence of the supergravity
background allows string interactions even at small radii, using
time dependent perturbation theory in an effective quantum mechanics.
We believe this approach will be useful in future studies of small-radius
Brandenberger-Vafa cosmology.

In conclusion, then, we have argued that it seems unlikely that string
gas cosmology in totally compact spaces provides an explanation for the
number of macroscopic dimensions, especially when our work is taken in
concert with \cite{Cleaver:1995bw,Easther:2003dd,Easther:2004sd}.  
Nonetheless, the Brandenberger-Vafa proposal
is a very natural initial condition for cosmology from the point of view
of string theory, so it is worthy of 
continued study.  We have in fact given some directions which we believe
would be interesting for future work.

\begin{acknowledgments}
We would like to acknowledge comments from R. Brandenberger and
M. Sheikh-Jabbari and useful 
conversations with J. Polchinski and especially P. Kraus.  The work of
ARF has been supported by a John A. McCone Fellowship in Theoretical Physics
at the California Institute of Technology.

\end{acknowledgments}

\appendix
\section{Quantum Radii}\label{a:qr}

In this appendix, we would like to sketch another
approach to resolving issues of energy conservation in completely 
compact spaces based on elementary considerations of quantum gravity.
In quantum gravity, we should really consider the metric to be some 
quantum variable, so that the radii of the compact dimensions should not be
fixed to the classical cosmological trajectories.  Rather, there should be
a wave function $\Psi(\mu,\nu)$ giving the probability density for different
values of $\mu,\nu$, which should be peaked at the classical trajectory.  
In fact, in the perturbative string regime, we expect that these are sharp
peaks because the Planck mass is very large (even compared to the string 
scale).  (We expect that $\Psi$ is most simply written in terms of $\mu,\nu$
because those variables have canonical kinetic terms.)

Then the total state of the system (at fixed time) should be described as 
$\ket{ \{\textnormal{matter}\},\mu,\nu}$, where the matter part describes
whatever string might be currently in the universe.  For simplicity, all but
two of the strings will be taken to be spectators (as is usual in perturbation
theory), and they will give only a factor of unity.  
Therefore, the total amplitude 
$\mathcal{A}=\inner{\{\textnormal{matter}\},\mu,\nu}{\{\textnormal{matter}\},
\mu,\nu}$ should be
\be\label{qgamp1}
\mathcal{A}=\int d\mu\, d\nu |\Psi(\mu,\nu)|^2 
\hat\mathcal{A}\delta\left(\sum E(\mu,\nu)
\right)\ .\ee
Here, $\hat\mathcal{A}$ is a reduced string
amplitude; just the amplitude from string perturbation theory with the energy
conserving delta function extracted.  
We will take the wavefunction to factorize on $\mu,\nu$.  To simplify notation,
we will henceforth let $E$ be the sum over energies (with outgoing energies
taken to be negative).  As an aside, we are ignoring any complications to 
the measure caused by our constraining sets of dimensions to have the same
radii.

Then suppose we have the simple case described above for winding mode 
annihilation.  If both the winding and momentum directions have the same
radius (ie, both are among the large dimensions represented by $\mu$), 
we can evaluate the integral easily to get 
\be\label{qgamp2}
\mathcal{A}=\frac{1}{4\sqrt{nw}}
\left|\Psi\left(\frac{1}{2}\ln\frac{n}{w}\right)\right|^2
\hat\mathcal{A}\ .\ee
Here we have integrated over the $\nu$ dependence of the wavefunction.
Other examples proceed in the same way, with a slight complication if the
energy depends on both $\mu$ and $\nu$.  In that case, 
we will find $\mu$ as a function of $\nu$ by
integrating over $\mu$ first.  Then generally we will have
\be\label{qgamp3}
\mathcal{A}=\int\, d\nu \left|\Psi(\b\mu,\nu)\right|^2 \left(
\frac{\del E}{\del \mu}(\b\mu,\nu)\right)^{-1} 
\hat\mathcal{A}(\b\mu,\nu)\ ,\ee
where $\bar\mu$ is the value of $\mu$ that satisfies conservation of
energy as a function of $\nu$.  
If the wavefunction is sharply peaked in $\nu$, then 
\be\label{qgamp4}
\mathcal{A}\simeq \left|\Psi(\bar\mu)\right|^2 \left(\frac{\del E}{\del \mu}
(\bar\mu)\right)^{-1} 
\hat\mathcal{A}(\bar\mu)\ ,\ee
where $\Psi$ now depends only on $\mu$ and everything else (including
$\bar\mu$) are evaluated at the peak (classical) value of $\nu$.

The interaction probability 
should be given by the normalized square of the the amplitude, as usual.  
To get the rate, however, we must divide the probability by amount of time
over which the interaction has been allowed to take place.  In perturbative
string theory, as in perturbative field theory, we have essentially assumed
that strings have been allowed to interact over some long period of time.
Usually, the square of the amplitude has two energy conserving delta functions,
and one of those delta functions, evaluated at zero, becomes the interaction
time.  In our analysis, the radial wavefunction replaces the energy
conserving delta function and should also represent the interaction time.

We can understand this fact by considering the classical limit of our
amplitude.  If we turn off quantum gravity (perhaps by sending the 
dimensionally reduced dilaton $\psi\to -\infty$), then we should have
$|\Psi(\mu,\nu)|^2=\delta(\mu-\hat\mu)\delta(\nu-\hat\nu)$, 
where $\hat\mu,\hat\nu$
are the classical values.  Then we obviously get the usual amplitude.
However, we can also integrate over $\mu,\nu$, so the amplitude is as
in (\ref{qgamp4}).  Then we note that 
\be\label{backtodelta}
\left|\Psi(\bar\mu)\right|^2 \left(\frac{\del E}{\del \mu}
(\bar\mu)\right)^{-1} = \delta(\bar\mu-\hat\mu) \left(\frac{\del E}{\del \mu}
(\bar\mu)\right)^{-1} =\delta(E(\hat\mu))\ .\ee
Therefore, when we square the amplitude, we have two factors of
$|\Psi|^2/(\del E/\del\mu)$, one of which should be evaluated at the classical
values $\hat\mu,\hat\nu$ and divided out to give the rate.

Once we average over incoming polarizations and sum over outgoing states,
we end up with a rate
\be\label{qgrate}
\Gamma = \mathcal{N}\sum_{\mathbf{s}} \sum_{\vec{n}}
\left|\Psi(\bar\mu,\hat\nu)\right|^2 \left(\frac{\del E}{\del \mu}
(\bar\mu,\hat\nu)\right)^{-1} 
\frac{|\hat\mathcal{A}
(\vec{w}_i,\vec{n}_i,\vec{n}_o,\bar\mu,\hat\nu)|^2}{(2E_{1})\cdots
(2E_{4})}
\ .\ee
Here the notation is as in equation (\ref{totalrate1}), and we have
specialized to the 2-string annihilation or scattering rate.

What is missing from this prescription is a wavefunction for the radii,
as well as its dynamics.  One particularly important question is how each
string interaction affects the wavefunction; that is, does the string
interaction ``collapse'' the wavefunction or is there some more complicated
entanglement or decoherence process?  Another point concerns the quantum
mechanical evolution of the wavefunction.  For example, 
if we choose to model the wavefunction
as a gaussian, how does the width of the gaussian evolve with time?

\bibliography{stringgas}

\end{document}